\documentclass[twocolumn,showpacs,showkeys,amsmath,amssymb,superscriptaddress,floatfix,nofootinbib]{revtex4}

\usepackage{graphicx}
\usepackage{bm} 
\usepackage{subfigure}

\makeatletter

\begin{document}
 
\title{Azimuthally-sensitive femtoscopy from RHIC to LHC in hydrodynamics with statistical hadronization%
\footnote{Supported in part by the Polish Ministry of Science and Higher Education, grants N202 153 32/4247, N202 034 32/0918, N202 249235, 
and by the U.S. NSF grant no. PHY-0653432.}}

\author{Adam Kisiel} 
\email{kisiel@if.pw.edu.pl}
\affiliation{Faculty of Physics, Warsaw University of Technology, PL-00661 Warsaw, Poland}
\affiliation{Department of Physics, Ohio State University, 
1040 Physics Research Building, 191 West Woodruff Ave., 
Columbus, OH 43210, USA}

\author{Wojciech Broniowski} 
\email{Wojciech.Broniowski@ifj.edu.pl} 
\affiliation{The H. Niewodnicza\'nski Institute of Nuclear Physics, Polish Academy of Sciences, PL-31342 Krak\'ow, Poland}
\affiliation{Institute of Physics, Jan Kochanowski University, PL-25406~Kielce, Poland} 

\author{Miko{\l}aj Chojnacki}
\email{Mikolaj.Chojnacki@ifj.edu.pl}
\affiliation{The H. Niewodnicza\'nski Institute of Nuclear Physics, Polish Academy of Sciences, PL-31342 Krak\'ow, Poland}

\author{Wojciech Florkowski} 
\email{Wojciech.Florkowski@ifj.edu.pl}
\affiliation{The H. Niewodnicza\'nski Institute of Nuclear Physics, Polish Academy of Sciences, PL-31342 Krak\'ow, Poland}
\affiliation{Institute of Physics, Jan Kochanowski University, PL-25406~Kielce, Poland} 


\begin{abstract}
Azimuthally-sensitive femtoscopy for heavy-ion collisions at RHIC and
LHC is explored within the approach consisting of the hydrodynamics of
perfect fluid followed by statistical hadronization. It is found that for the
RHIC initial conditions the very same framework that reproduces the
standard soft observables (including the transverse-momentum spectra,
the elliptic flow, and the azimuthally-averaged HBT radii) leads to
a proper description of the azimuthally-sensitive femtoscopic
observables - we find that the azimuthal variation of the {\em side}
and {\em out} HBT radii is very well reproduced for all centralities,
while the {\em out-side} correlation is somewhat too large for
non-central events. Concerning the dependence of the femtoscopic
parameters on $k_T$ we find that it is very well reproduced for the
{\em out} and {\em side} radii, and fairly well for the {\em long}
radius. The model is then extrapolated for the LHC energy. 
We predict the overall moderate growth of
the HBT radii and the decrease of their azimuthal  oscillations. Such
effects are naturally caused by longer evolution times.   In addition,
we discuss in detail the space-time patterns of particle emission. We
show that they are quite complex and argue that the overall shape seen
by the femtoscopic methods cannot be easily disentangled on the basis of
simple-minded arguments.  
\end{abstract}



\maketitle 


\section{Introduction}

Femtoscopy provides detailed information on the dynamics of systems
formed in relativistic heavy-ion collisions (for a review and
literature see \cite{Lisa:2005dd}). In particular, the
Hanbury-Brown-Twiss (HBT)
\cite{HanburyBrown:1954wr,HanburyBrown:1956pf,Goldhaber:1960sf,Gyulassy:1979yi}
intensity interferometry exhibits sensitivity to shape and flow of the
medium at freeze-out, which in turn reflects the conditions throughout
the evolution, from the formation till the cease of
interactions. Together with other observed quantities, such as the
ratios of particle abundances, momentum spectra, the elliptic flow
coefficient, or other correlation data, femtoscopy contributes to the
growing precise knowledge of the dynamics of the system. Azimuthally
sensitive HBT interferometry (azHBT)
\cite{Voloshin:azHBT,Lisa:2000xj}, which is the subject of this paper,
brings in information on the dependence of shape and flow on the
azimuthal angle $\phi$. This information is complementary to the data
on the transverse-momentum elliptic coefficient $v_2$
\cite{Ollitrault:1992bk}, which is a measure of the azimuthal
asymmetry of the flow. Thus, a simultaneous description of azHBT and $v_2$
verifies consistency of the theoretical description.

In this work we use what we call the {\em standard approach},
consisting of ideal-fluid hydrodynamics followed by statistical
hadronization. Numerous calculations have been done in this framework,
with the common difficulty \cite{Heinz:2002un} of simultaneously
describing femtoscopy and other signatures of the data. More
precisely, the {\em RHIC HBT puzzle}
\cite{Heinz:2002un,Hirano:2004ta,Lisa:2005dd,Huovinen:2006jp} refers
to  problems in reconciling the large value of the elliptic flow
coefficient, $v_2$, with the HBT interferometry in calculations based
on hydrodynamics
\cite{Heinz:2001xi,Hirano:2001yi,Hirano:2002hv,Zschiesche:2001dx,Socolowski:2004hw}.
In the standard approach the description of $v_2$ requires longer
hydrodynamic evolution times, while the HBT radii, sensitive to the
size and lifetime of the system, are properly reproduced when the
lifetime of the system is short. The two requirements are in conflict. 
Due to the obstacle of the RHIC HBT puzzle, only few calculations or
physically motivated parametrization of the azHBT quantities have
been made
\cite{Heinz:2002sq,Tomasik:2004bn,Tomasik:2005ny,Humanic:2005ye,Frodermann:2007ab,Csanad:2008af,Retiere:2003kf}.
Moreover, in our view the comparison to data of the azimuthal
dependence of the correlation radii makes sense only when the values
averaged over $\phi$ are also properly reproduced.  

Recently, we have accomplished a successful uniform description of
soft observables at RHIC, including the HBT radii, within the standard
approach \cite{Broniowski:2008vp}. The essential ingredients of this
analysis are the Gaussian initial condition for hydrodynamics, early
start of the evolution, the state-of-the art equation of state with
smooth crossover, and the use of {\tt THERMINATOR}
\cite{Kisiel:2005hn} with all resonances from {\tt SHARE}
\cite{Torrieri:2004zz} incorporated to carry out the statistical
hadronization at the freeze-out surface of temperature $145$~MeV. The
interplay of these elements resulted in a simultaneous description of
the transverse-momentum spectra of pions, kaons and protons, the $v_2$,
and the HBT correlation radii of pions, including the basic azHBT
signatures. The agreement of this quality cannot be achieved with the
initial condition obtained from the typically used Glauber models. In that
case the HBT radii, in particular the ratio of {\em out} to {\em side}
radii, is reproduced at the level of 20\% only \cite{Chojnacki:2007rq}. The
reader is referred to Ref.~\cite{Broniowski:2008vp} for the details.  

In this paper we present a systematic study of the azimuthally
sensitive HBT radii in the model of Ref.~\cite{Broniowski:2008vp},
including the dependence on centrality and the transverse momentum of
the pair, $k_T$. We find that the azimuthal variation of the {\em
side} and {\em out} radii is properly reproduced for all centralities,
while in the case of the {\em out-side} correlation it is somewhat too
large for non-central events. Concerning the dependence of the
femtoscopic parameters on $k_T$, it is very well reproduced for the
{\em out} and {\em side} radii, and fairly well for the {\em long}
radius. We stress that the present study involves no parametric
freedom, as all parameters have been fixed in the global fits of
Ref.~\cite{Broniowski:2008vp}. 

Next, we extrapolate the model to the LHC energy, where predictions
for the HBT and azHBT quantities is made. In addition, we present a
detailed analysis of the space-time patterns of the pion emission. We
argue that this emission, consisting of surface and volume parts,
exhibits a rather complex behavior. We argue that the determination of
the overall shape of the source seen by the femtoscopic methods is not
straightforward and requires detailed simulation, such as the one
performed in this work. 

We use $c=1$ throughout the paper. The label RHIC denotes the AuAu
collisions at the highest RHIC energy of $\sqrt{s_{NN}}=200$~GeV,
while LHC corresponds to the PbPb collisions at $\sqrt{s_{NN}}=5500$~GeV.

\section{The framework}

In this section we describe the essential elements of our method to
the extent they are necessary for the comprehensive presentation of
the new results. More details concerning the hydrodynamics can be
found in
Refs.~\cite{Chojnacki:2006tv,Chojnacki:2007jc,Broniowski:2008vp},
while the method used for femtoscopic calculations has been presented
in great detail in Ref.~\cite{Kisiel:2006is}.  

\subsection{Initial condition}

As reported in Ref.~\cite{Broniowski:2008vp}, the use of the initial condition for hydrodynamics of
the Gaussian form,
\begin{eqnarray}
n(x,y)=\exp \left ( -\frac{x^2}{2a^2} -\frac{y^2}{2 b^2} \right ),
\label{profile}
\end{eqnarray} 
where $x$ and $y$ denote the transverse coordinates, leads to a much
better  uniform description of the data for the $p_T$-spectra, $v_2$,
and the pionic HBT radii compared to the use of the standard initial
condition from the Glauber model.  

\begin{figure}[tb]
\begin{center}
\includegraphics[angle=0,width=0.48 \textwidth]{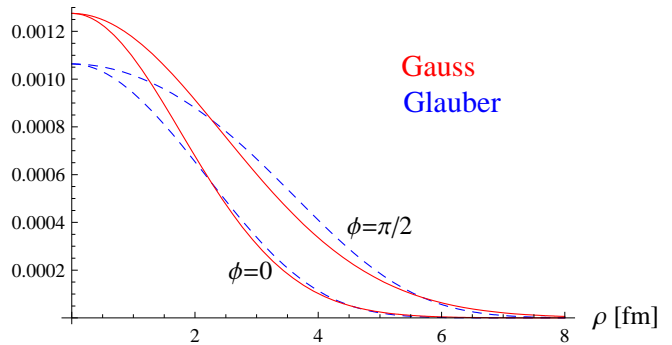}
\end{center}
\vspace{-6.5mm}
\caption{(Color online) In-plane and out-of-plane sections of the
two-dimensional  energy-density profiles for $c=30-40\%$ obtained from
the mixed Glauber model described in the text (dashed lines) and the
Gaussian  parametrization (\ref{profile}) used in this work (solid
lines). Both profiles are normalized to unity,  and in addition
$\langle x^2 \rangle=a^2$, $\langle y^2 \rangle=b^2$.  
\label{fig:shape}}
\end{figure}

The width parameters $a$ and $b$ depend on centrality. In order to use
realistic  values we run the {\tt GLISSANDO} \cite{Broniowski:2007nz}
Glauber Monte Carlo simulations which include the eccentricity
fluctuations \cite{Andrade:2006yh,Hama:2007dq}. Then we match $a^2$
and $b^2$ to reproduce the values $\langle x^2 \rangle$ and $\langle
y^2 \rangle$ from the {\tt GLISSANDO} profiles. Thus, by construction,
the spatial rms radii of the initial condition and its eccentricity is
the same as from the Glauber calculation. Nevertheless, the shape is
not the same, as is evident from Fig.~\ref{fig:shape}. The Gaussian
profiles are sharper near the origin, which results in a faster
buildup of the Hubble-like flow in the hydrodynamical stage.  
Admittedly, the initial density and flow profiles should eventually be
obtained from the early dynamics, such as the Color Glass Condensate
theory \cite{McLerran:1993ni,McLerran:1993ka,Kharzeev:2001yq}. In
practice, however, modeling of the partonic stage carries uncertainty
in its parameters. In addition, other effects in the early dynamics
are present, see e.g. \cite{Mrowczynski:1993qm,Muller:2007rs}, making
precise profile calculations difficult. Thus the use of a simple
parametrization of the initial profile is a profitable and practical
approach, while the need remains for its detailed dynamical
justification. 

\begin{table*}[tb]
\caption{Shape parameters $a$ and $b$ of Eq.~(\ref{profile}) for various centrality classes 
obtained by matching  $\langle x^2 \rangle$ and $\langle y^2 \rangle$ to 
{\tt GLISSANDO} simulations, the variable-axes eccentricity
$\epsilon^\ast$, and the chosen central temperature $T_i$. 
\label{tab:ab}}
\begin{tabular}{|r|rrrrrrrrr|}
\hline
$c$ [\%] 		&  0- 5 &  5-10 & 10-20 & 20-30 & 30-40 & 40-50 & 50-60 & 60-70 & 70-80 \\
\hline
 & \multicolumn{9}{|c|}{RHIC}\\
$a$ [fm]		& 2.70  & 2.54  & 2.38  & 2.00  & 1.77  & 1.58  & 1.40  & 1.22  & 1.04  \\
$b$ [fm] 		& 2.93  & 2.85  & 2.74  & 2.59  & 2.45  & 2.31  & 2.16  & 2.02  & 1.85  \\
$\epsilon^\ast$ & 0.08  & 0.12  & 0.18  & 0.25  & 0.31  & 0.36  & 0.41  & 0.46  & 0.52  \\
$T_i$ [MeV]     & 500   & 491   & 476   & 460   & 429   & 390   & 344   & 303   & 261   \\
\hline
 & \multicolumn{9}{|c|}{LHC}\\
$a$ [fm] 		& 2.65  & 2.47  & 2.22  & 1.95  & 1.73  & 1.56  & 1.40  & 1.24  & 1.06  \\
$b$ [fm] 		& 2.89  & 2.80  & 2.69  & 2.55  & 2.41  & 2.28  & 2.15  & 2.02  & 1.85  \\
$\epsilon^\ast$ & 0.09  & 0.12  & 0.19  & 0.26  & 0.32  & 0.36  & 0.40  & 0.45  & 0.51  \\
$T_i$ [MeV] 	& 768   & 768   & 768   & 768   & 768   & 768   & 768   & 768   & 768   \\
\hline
\end{tabular}
\end{table*}

The Glauber calculations, needed to obtain the $a$ and $b$ parameters,
correspond to the mixed model \cite{Kharzeev:2000ph}, where the number
pf produced particles is proportional to  $(1-\alpha)N_w/2 + \alpha
N_{\rm bin}$, with $N_w$ and $N_{\rm bin}$ denoting the number of
wounded nucleons \cite{Bialas:1976ed} and binary collisions,
respectively. The parameter $\alpha=0.145$  for RHIC
\cite{Back:2001xy,Back:2004dy} and is set by us to $0.2$ for LHC. The
inelastic nucleon cross section is 42~mb for RHIC and 63~mb for LHC
\cite{Yao:2006px}. The simulations incorporate the fluctuations of
orientation of the fireball (the variable-axes geometry), which result
in increased eccentricity compared to the fixed-axes geometry
\cite{Broniowski:2007ft}. Finally, the expulsion distance of 0.4~fm is
used in the generation of the nuclear distributions, and the
source-dispersion parameter of 0.7~fm is used. This parameter
describes the random displacement of the source from the center of the
wounded nucleon or the binary-collision position
\cite{Broniowski:2007nz}. The values of the $a$ and $b$ parameters for
various centralities and the corresponding eccentricity parameters 
\begin{eqnarray}
\epsilon^\ast=\frac{b^2-a^2}{a^2+b^2}, \label{eps}
\end{eqnarray}
are collected in Table~\ref{tab:ab}. 

The energy-density profile (\ref{profile}) determines the initial
temperature profile via the equation of state
\cite{Chojnacki:2007jc}. The initial central temperature, $T_i$, is a
model parameter dependent on centrality. For RHIC calculations it is
adjusted to reproduce the total particle multiplicity. For the LHC we 
use for simplicity a common value of $768$~MeV at $\tau_{0}=0.25$~{\rm
fm} which corresponds to $500$~MeV at $\tau=1$~fm. When the LHC
multiplicity data are available in the future, this parameter will be
tuned more realistically.  

\subsection{Hydrodynamics}

The hydrodynamic equations used in this work were described in detail
in Refs.~\cite{Chojnacki:2006tv,Chojnacki:2007jc}. We use inviscid
(ideal-fluid), baryon-free, boost-invariant hydrodynamics. The
equations are written in terms of the velocity of sound, $c_s$, whose
temperature dependence encodes the full information on the equation of
state of the system. Importantly, we incorporate the known features of
$c_s(T)$, which at high temperatures are given by the lattice QCD
calculations, at low $T$ follow from the hadron gas including all
resonances, while in between an interpolation is used. Importantly, no
phase transition, but a smooth cross-over is built in, in accordance
to the present knowledge of the thermodynamics of QCD at zero baryon
chemical potential. The plot of the resulting $c_s(T)$ can be found in
Ref.~\cite{Broniowski:2008vp}. 

The initial proper time for the start of hydrodynamics is fixed to
have the value 
\begin{eqnarray}
\tau_0=0.25{\rm ~fm}
\end{eqnarray}
for all centralities, both for RHIC and LHC.
This early start of hydrodynamics allows for a fast generation of transverse flow.%
\footnote{The ignition of hydrodynamics may be delayed to later times, about 1~fm, if it is preceded
with particle free-streaming starting at 0.25~fm followed by the Landau matching \cite{Broniowski:2008vp}. 
This mechanism generates some initial flow (transverse and elliptic) for hydrodynamics. The resulting 
freeze-out hypersurfaces are very close to those used in this work.}

\subsection{Freeze-out}

\begin{figure*}[tb]
\begin{center}
\includegraphics[angle=0,width=0.95 \textwidth]{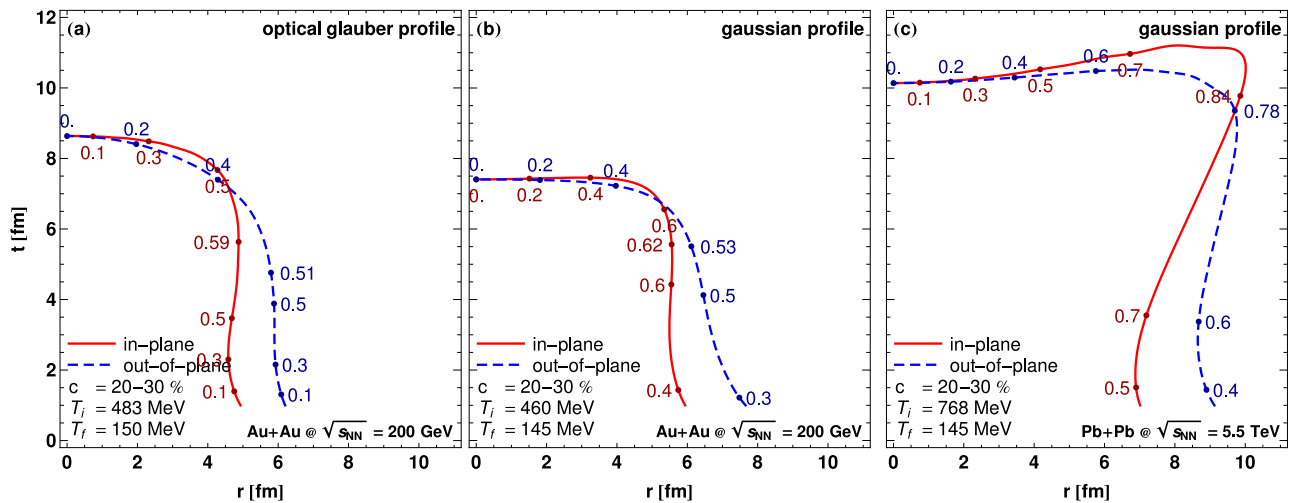}
\end{center}
\vspace{-6.5mm}
\caption{(Color online) Sections of the freeze-out hypersurfaces for
the Glauber initial condition used in Ref.~ \cite{Chojnacki:2007rq}
(left) and for the Gaussian initial condition (middle) for centrality
20-30\% for RHIC, and for the Gaussian initial condition  for
centrality 20-30\% for LHC (right). The dots with numbers indicate the
values of the transverse velocity at freeze-out. 
\label{fig:hs}}
\end{figure*}

The hydrodynamic evolution proceeds until the freeze-out occurs, where the
assumed condition for the {\em universal} freeze-out temperature is
$T_f=145$~MeV. This value is somewhat lower than in several fits of
the chemical freeze-out
\cite{BraunMunzinger:2001ip,Broniowski:2001we,Broniowski:2001uk},
however, it agrees with the recently made global fits to particle
transverse momentum spectra of
Ref.~\cite{Rafelski:2004dp,Prorok:2007xp}, where the value around
145~MeV was obtained for the kinetic freeze-out
temperature.\footnote{The use of this lower freeze-out temperature
needs the introduction of the strangeness inequilibrium factors
$\gamma_s$ in order to reproduce the abundances of strange particles
\cite{Rafelski:2004dp}.}  

The freeze-out hypersurfaces are compared in Fig.~\ref{fig:hs}. We use the centrality 20-30\% as 
an example, as for other cases the results are qualitatively similar. 
The left and middle panels show the role of the initial condition used for the RHIC analysis.
For the Glauber initial condition used in
Ref.~\cite{Chojnacki:2007rq} (left) the freeze-out hypersurface is smaller in transverse size but longer in 
the life-time than for the case of the Gaussian initial
condition (middle) used in the present work. Parameters in both
calculations were optimized to reproduce the yields, the $p_T$ spectra, and $v_2$. The fact 
that for the Gaussian initial profile the source has a significantly
larger transverse size and shorter emission time is a crucial feature
for the successful description of the HBT data. 

The freeze-out hypersurfaces have the {\em volume} emission parts
(with time-like normal vectors, i.e. running approximately flat in
Fig.~\ref{fig:hs}), which are similar to the blast-wave
parametrization \cite{Schnedermann:1993ws}. The lifetime is short:
about 9~fm for central \mbox{(c = 0-5\%)} and 7~fm for mid-peripheral
(c = 20-30\%) collisions at RHIC. However, they also contain the {\em surface}
emission parts (with space-like normal vectors, i.e. running more-or-less vertical
in Fig.~\ref{fig:hs}), absent in the usual blast-wave
parametrizations. Nevertheless, there is no problem with the
non-causal surface emission
\cite{Bugaev:1996zq,Anderlik:1998et,Borysova:2005ng}, which is
negligible, since less than 0.5\% of particles are emitted back into
the hydrodynamic region. This very small fraction follows from the
large transverse flow velocity at large radii, as indicated by the
labels in Fig.~\ref{fig:hs}. This large flow carries the particles
outward. 
In addition, for the RHIC case 
the hypersurfaces are not bent back at low times, as
occurs in some hydrodynamic calculations. Quantitatively, we have
found that about half of the produced particles comes from the volume
emission part and about half from the surface emission part. The
relevance of the surface emission is also stressed in
Refs.~\cite{Sinyukov:2006dw,Gyulassy:2007zz}. 

The right panel in Fig.~\ref{fig:hs} shows the freeze-out hypersurface
for the LHC extrapolation for the Gaussian initial profile and the
same centrality.  We note a larger transverse size and emission times
compared to the RHIC case of the middle panel.  The surface-emission
parts of the curves are bent back, but at the same time the expansion
velocity is larger, such that the backward emission problem is again
irrelevant.  

{\tt THERMINATOR} is used to carry out the statistical hadronization
at the freeze-out hypersurface according to the Cooper-Frye
formulation \cite{Cooper:1974mv}. According to the assumed
single-freeze-out approximation, identifying the kinetic and chemical
freeze-out temperatures, rescattering process after freeze-out are
neglected. We have checked that the collision rate after freeze-out is
moderate for the hypersurfaces applied in this work. We estimate it by
considering a pion straight-line trajectory and counting the number of
encounters with other particles closer than the distance corresponding
to the pion-hadron cross section. The average number of these
trajectory crossings is about 1.5-1.7 per pion. This shows that the
single-freeze-out approximation \cite{Broniowski:2001we} works
reasonably well for the present case. At a more detailed level, one
could use hadronic afterburners to model the elastic collisions 
\cite{Teaney:2000cw,Nonaka:2006yn,Hirano:2007xd}, or attempt the
hydro-kinetic approach implemented in \cite{Amelin:2006qe}. 

\subsection{Model parameters}

For the reader's convenience, we list here again all the model
parameters. The model has the total of 5 parameters. Parameters
dependent on the centrality classes are the Gaussian widths, $a$, $b$,
and the initial central temperature, $T_i$. The widths are fixed with
{\tt GLISSANDO} \cite{Broniowski:2007nz}, while $T_i$ is adjusted in
order to reproduce the particle multiplicities. The parameters
independent of centrality are the starting proper time of
hydrodynamics, $\tau_0$, and the universal freeze-out temperature,
$T_f$.

\subsection{Two-particle femtoscopy}

The method of dealing with the HBT quantities has been thoroughly described
in Ref.~\cite{Kisiel:2006is}, where the reader is referred to
for details. Here we only note that our technique follows exactly the
experimental procedure of extracting the femtoscopic quantities,
including the passage to the LCMS (local co-moving system) frame, the
application of the two-particle method, and carrying out the
Bowler-Sinyukov procedure \cite{Bowler:1991vx,Sinyukov:1998fc} to
incorporate the Coulomb corrections. 

\begin{figure*}[tb]
\begin{center}
\includegraphics[angle=0,width=0.9 \textwidth]{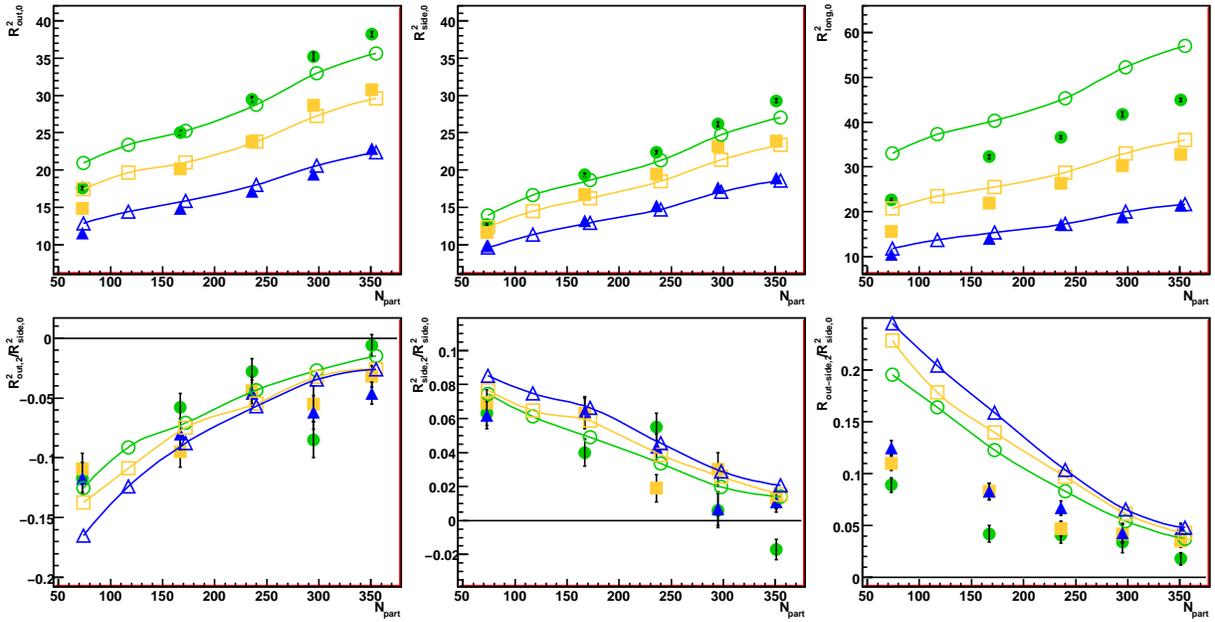}
\end{center}
\vspace{-6.5mm}
\caption{(Color online) 
Results for the RHIC HBT radii and their azimuthal oscillations. For
each value of $N_{\rm part}$ on the horizontal axis we plot the
experimental points (filled symbols) and the model results (empty
symbols). The points from top to bottom at each plot correspond to
$k_T$ contained in the bins 0.15-0.25~GeV (circles),
0.25-0.35~GeV  (squares), and 0.35-0.6~GeV(triangles). The top panels
show $R^{2}_{\rm out,0}$, $R^{2}_{\rm side,0}$, and $R^{2}_{\rm
long,0}$, the bottom panels the magnitude of the allowed
oscillations divided conventionally by $R_{\rm side,0}^2$. 
Data from Ref.~\cite{Adams:2003ra}.
\label{fig:resultsrhic}}
\end{figure*}

We use the standard Pratt-Bertsch parametrization \cite{Pratt:1986cc,Bertsch:1989vn}.
For the case without azimuthal symmetry the correlation function is fitted to the formula 
including the cross terms between the HBT radii \cite{Heinz:2002au}:
\begin{eqnarray}
&& C(q_{\rm out},q_{\rm side},q_{\rm long})  =  1+\lambda \exp(-R_{\rm out}^{2} q_{\rm out}^2 -
R_{\rm side}^2 q_{\rm side}^2 \nonumber \\
&&- R_{\rm long}^2 q_{\rm long}^2 - R_{\rm out-side}q_{\rm out}q_{\rm side} - R_{\rm out-long} q_{\rm out} q_{\rm long} 
\nonumber \\ &&- R_{\rm side-long} q_{\rm side} q_{\rm long})
\end{eqnarray}
The basic quantities of the 
azimuthally sensitive HBT analysis are the second-order Fourier coefficients defined as 
\begin{eqnarray}
R_{\rm out}^2(\phi) &=& R_{\rm out,0}^2 + 2R_{\rm out,2}^2 \cos(2\phi), \nonumber \\
R_{\rm side}^2(\phi) &=& R_{\rm side,0}^2 + 2R_{\rm side,2}^2 \cos(2\phi), \nonumber  \\
R_{\rm long}^2(\phi) &=& R_{\rm long,0}^2 + 2R_{\rm long,2}^2 \cos(2\phi), \nonumber  \\
R_{\rm out-side}(\phi) &=& 2R_{\rm out-side,2} \sin(2\phi),  \nonumber \\
R_{\rm out-long}(\phi) &=& 2R_{\rm out-long,2} \cos(2\phi), \nonumber  \\
R_{\rm side-long}(\phi) &=& 2R_{\rm side-long,2} \sin(2\phi). 
\label{eq:oscilations}
\end{eqnarray}

\section{Results}

\subsection{RHIC}

In Fig.~\ref{fig:resultsrhic} we present the summary of our results 
compared to the experimental data from the STAR AuAu collisions at
$\sqrt{s_{NN}}=200$~AGeV~\cite{Adams:2003ra}. The figure shows the
centrality and pair transverse momentum ($k_T$) dependence. For each
centrality, associated here with the number of participants $N_{\rm
part}$ on the horizontal axis, we plot the experimental points (filled
dots) and the model results (empty symbols). The points from top to
bottom correspond to $k_T$ contained in the bins of 0.15-0.25~GeV,
0.25-0.35~GeV, and 0.35-0.6~GeV. The top panels show the radii squared
averaged over the $\phi$ angle, from left to right, $R^{2}_{\rm
out,0}$, $R^{2}_{\rm side,0}$, and $R^{2}_{\rm long,0}$. The bottom
panels show the magnitude of the allowed oscillations divided by
$R_{\rm side,0}^2$, which is the adopted convention used in presenting
the experimental data.  

The values of the model points in the plots were obtained by first 
solving the hydrodynamic equations and running {\tt THERMINATOR} 
as described in the previous sections.
Then, according to our procedure, 144
separate correlation functions have been constructed (6 centrality bins
$\times$ 4 $k_T$ bins $\times$ 6 bins for the phi
angle). The $\phi$ dependence of the correlation function was obtained
in the following way: an angle between the $k_T$
direction and the reaction plane was determined and the pairs were grouped
into six bins: ($-\frac{\pi}{12}, \frac{\pi}{12}$), ($\frac{\pi}{12}, \frac{3\pi}{12}$),
($\frac{3\pi}{12}, \frac{5\pi}{12}$), ($\frac{5\pi}{12}, \frac{7\pi}{12}$), 
($\frac{7\pi}{12}, \frac{9\pi}{12}$), and ($\frac{9\pi}{12},\frac{11\pi}{12}$). 
Each of the six size parameters was then plotted vs. the reaction
plane orientation for a given centrality and $k_T$ bin. The
dependence was fit with the formula~(\ref{eq:oscilations}) including
only the allowed oscillations, and the fit values were corrected for
the finite bin size~\cite{Heinz:2002au}. 
By symmetry arguments, the oscillations in the {\em long}, {\em
side-long} and {\em out-long} components must vanish, as they do
within statistical error in our calculations. This is the reason why
they are not shown in the plot.  

\begin{figure*}[tb]
\begin{center}
\includegraphics[angle=0,width=0.7 \textwidth]{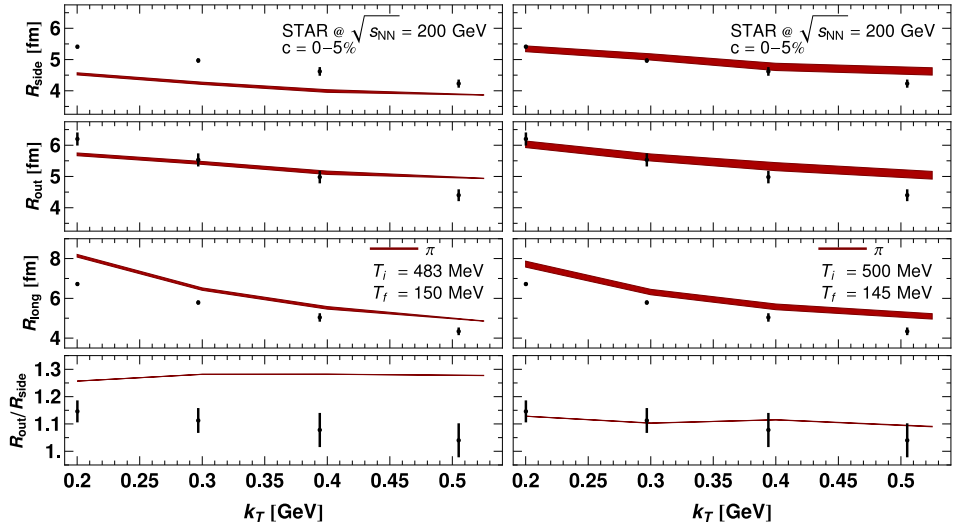}
\end{center}
\vspace{-6.5mm}
\caption{(Color online) Results for the angle-averaged {\rm side},
{\em out}, and {\em long} radii, and the ratio $R_{\rm out}/R_{\rm
side}$, for the Glauber initial condition (left) and for the Gaussian
initial condition (right). Data from Ref.~\cite{Adams:2003ra}. 
\label{fig:hbtcomp}}
\end{figure*}

Fig.~\ref{fig:resultsrhic} shows a very good agreement for $R_{\rm
out,0}$ and $R_{\rm side,0}$, where the model points are close the the
experiment at all centralities and $k_T$ bins. The agreement for
$R_{\rm long,0}^2$ is somewhat worse in the lowest $k_T$ bin, with the
model overshooting the data by up to 10\% for the values of the
radius itself. This may be due to the assumed exact boost invariance
in the model calculation. The oscillations  $R_{\rm side,2}^2$ and
$R_{\rm out,2}^2$ are in a very good agreement with the data. The
oscillations in the cross term $R_{\rm out-side}$ are about 50\% above
the data. This mismatch requires explanation.   

\subsection{Gaussian vs. Glauber initial condition}

\begin{figure*}[tb]
\begin{center}
\includegraphics[angle=0,width=0.99 \textwidth]{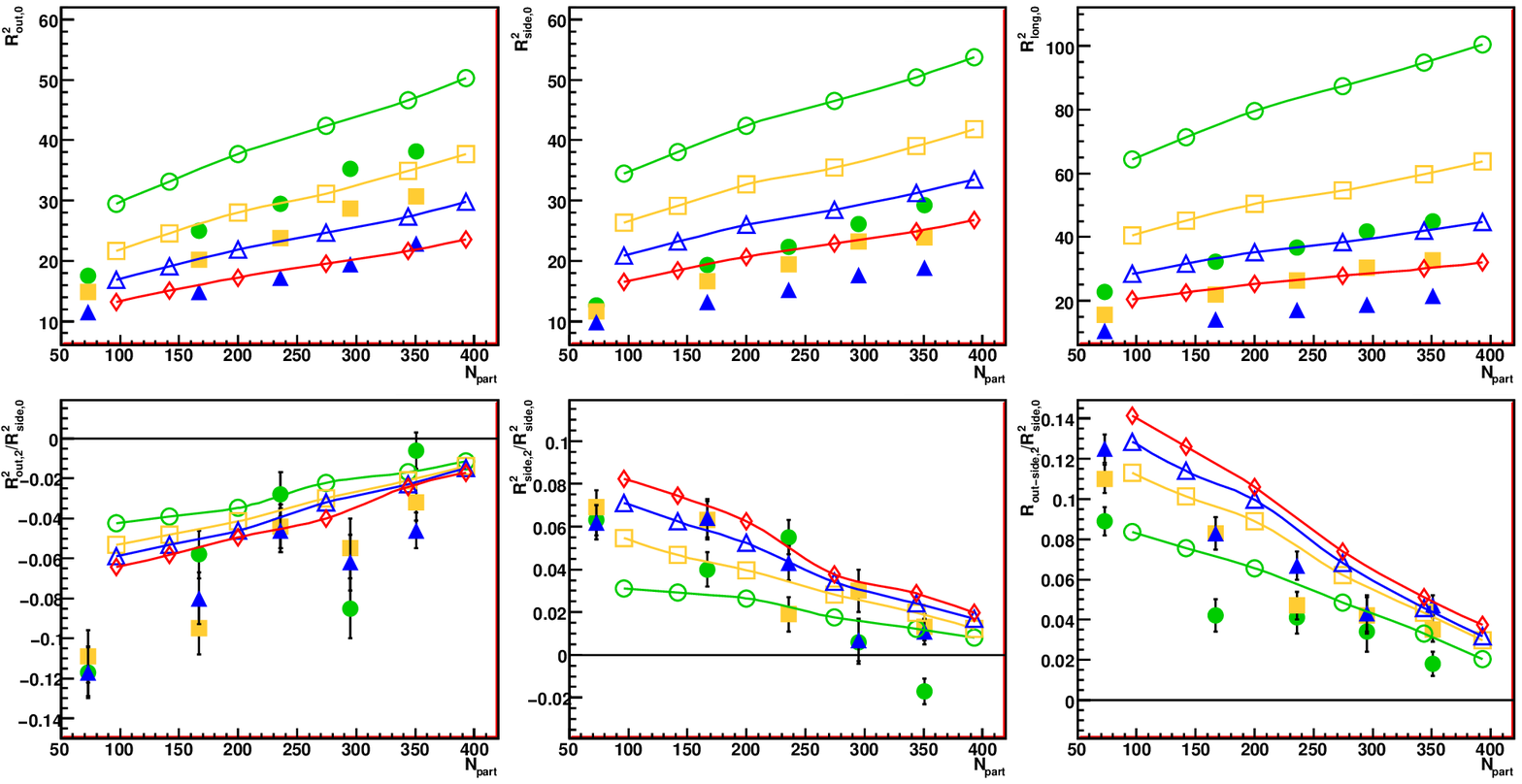}
\end{center}
\vspace{-6.5mm}
\caption{(Color online) Predictions for the LHC. Filled symbols are
RHIC data from STAR, shown for comparison (circles - $k_T$ in
0.15-0.25 GeV, squares - $k_T$ in 0.25-0.25 GeV, triangles - $k_T$ in
0.35-0.6 GeV). Open symbols are the results of our calculation for the
LHC energy with the initial temperature set to $768$~MeV. The last
$k_T$ bin has been divided into two: open triangles - $k_T$ in
0.35-0.45 GeV and open diamonds - $k_T$ in 0.45-0.6~GeV. Top panels
show sizes (squared) averaged over the $\phi$ angle.  Bottom panels
show magnitude of the allowed oscillations, divided by $R_{\rm
side}^2$. Data from Ref.~\cite{Adams:2003ra}.
\label{fig:resultslhc}}
\end{figure*}

In order to point out the relevance of the initial condition mentioned
in Ref.~\cite{Broniowski:2008vp}, we compare in Fig.~\ref{fig:hbtcomp}
the angle-averaged {\em side}, {\em out}, and {\em long} radii, and
the ratio $R_{\rm out}/R_{\rm side}$, obtained with the Glauber and
Gaussian initial conditions. We note a much better agreement with the
presently applied Gaussian initial condition. In particular, the ratio
$R_{\rm out}/R_{\rm side}$ coincides now with the experiment, while
with the Glauber profile it was about 20\% above the data and the
$k_T$-dependence was not reproduced. 

\subsection{Predictions for LHC}

In Fig.~\ref{fig:resultslhc} we show a plot analogous to
Fig.~\ref{fig:resultsrhic} but with calculations done for the LHC
energies, assuming the initial temperature of $768$~MeV. As expected,
the larger radii reflect the growth of the overall size of the system
at freeze-out. All three radii grow, but the $out$ radius seems 
to grow less, which is discussed in detail later. Also the azimuthal
oscillations relative to $R_{\rm side}^2$ are smaller compared to RHIC. Again, this is
expected: the initial asymmetry of the system at a given centrality is
quite similar at RHIC and at the LHC, since it is mainly driven by the
overlap geometry of the two colliding nuclei. The system then evolves from the out-of-plane extended
source towards the more spherical shape, and if the evolution time is long
enough, it may eventually overshoot and become in-plane
extended. Measurements at RHIC show that the source freezes out while
still in the out-of-plane shape. Our calculations for LHC show that the
evolution time to freeze-out is longer, but not long enough to produce
the overshoot. The smaller oscillations (relative to $R_{\rm side}^2$) are a consequence of the fact
that the system has been evolving for a longer time and effectively becomes more spherical
than at RHIC. 

\begin{figure*}[tb]
\begin{center}
\includegraphics[angle=0,width=0.7 \textwidth]{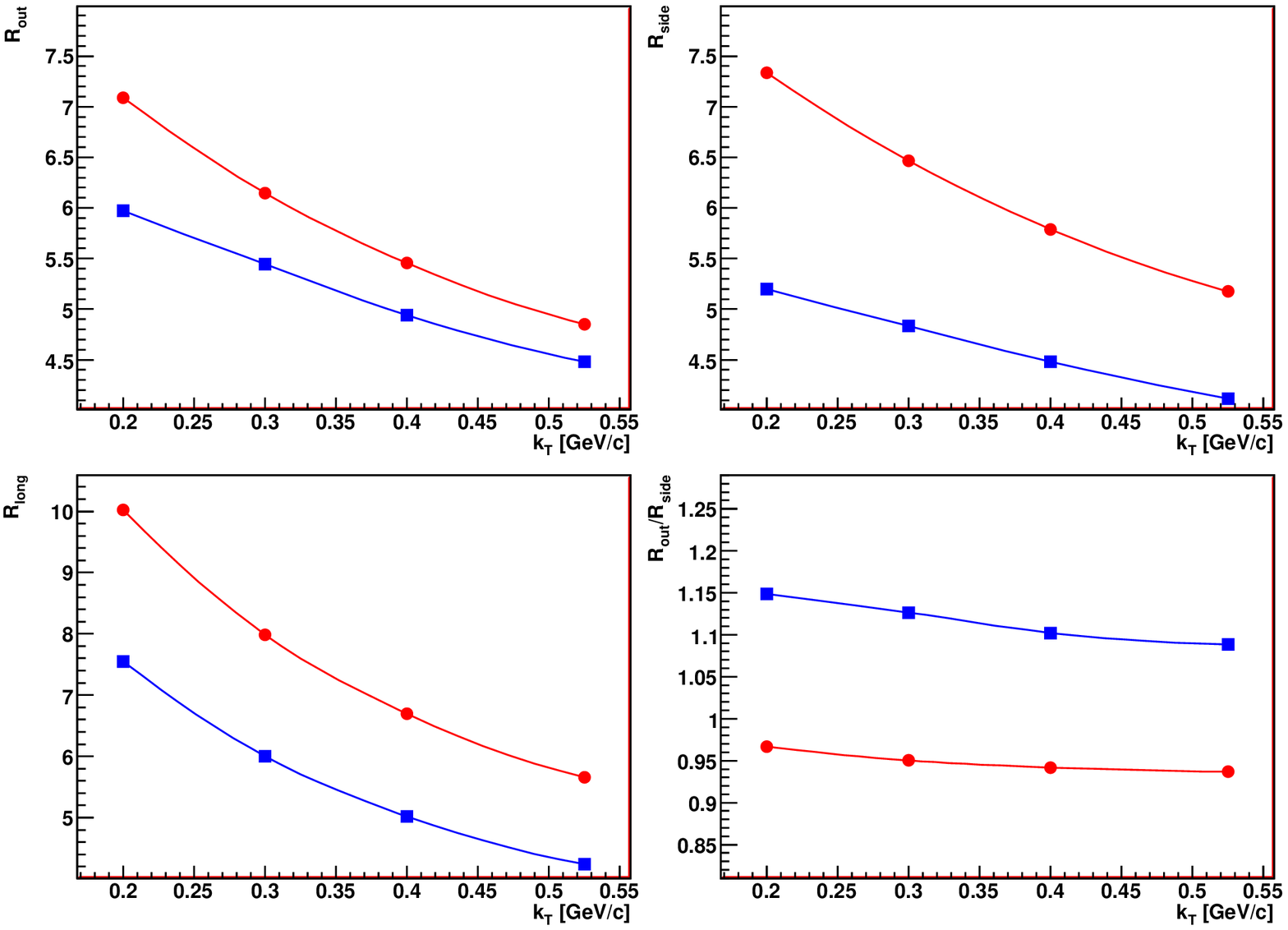}
\end{center}
\vspace{-6.5mm}
\caption{(Color online) Comparison of calculations for the top 5\%
central AuAu collisions at RHIC (squares) and PbPb collisions at LHC
(circles) energies. Lines are drawn to guide the eye. Top-left panel
shows {\em out} radius, top-right - {\em side} radius, bottom left - {\em long} radius,
bottom right - {\em out} over {\em side} radii ratio.
\label{fig:resultsrhictolhc}}
\end{figure*}

\begin{figure*}[tb]
\begin{center}
\includegraphics[angle=0,width=0.99 \textwidth]{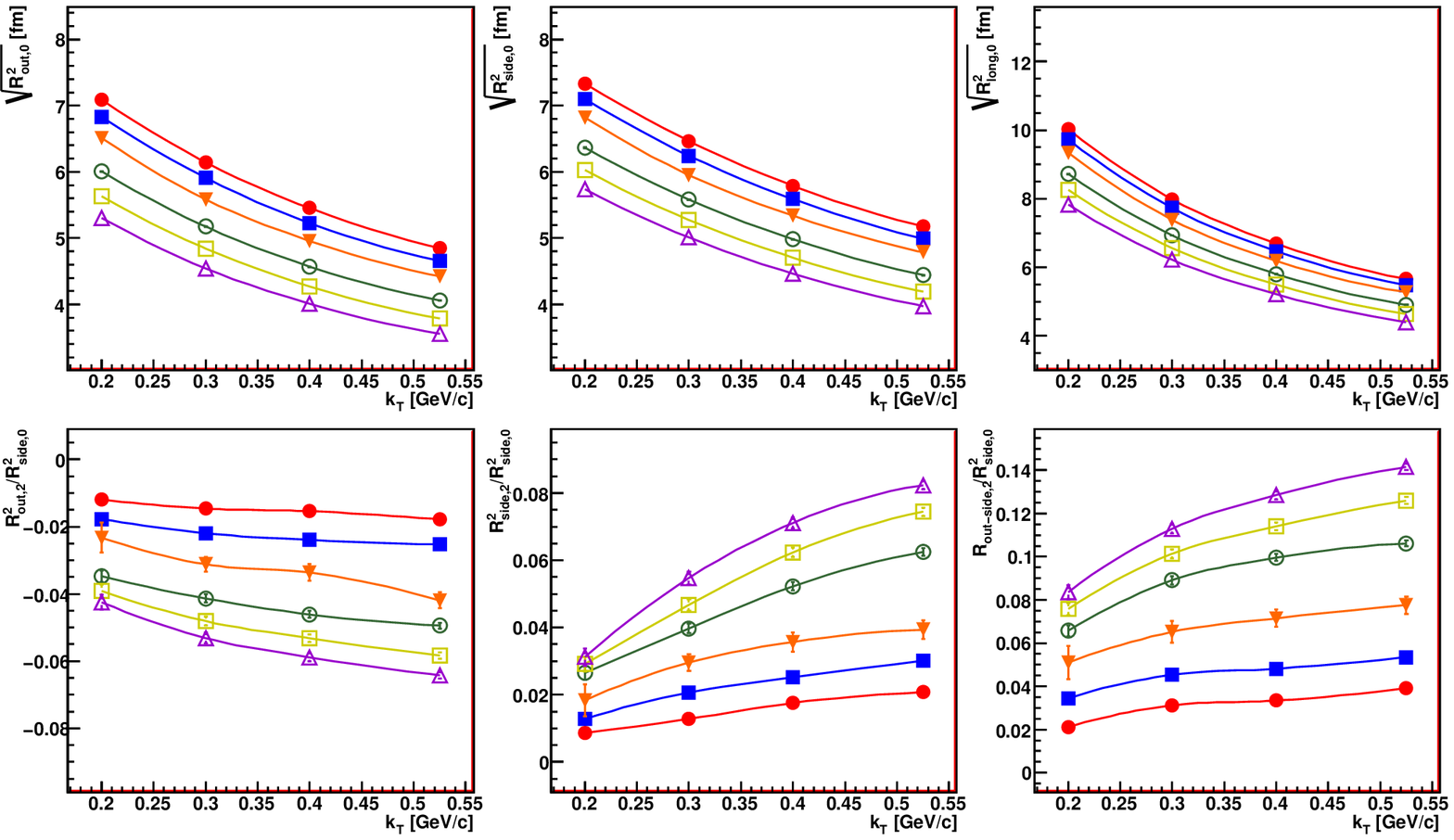}
\end{center}
\vspace{-6.5mm}
\caption{(Color online) Calculations for the LHC with the Gaussian initial conditions for 6
centralities (filled circles - 0-5\%, filled squares -
5-10\%, filled down triangles 10-20\%, open circles -
20-30\%, open squares - 30-40\%, open up triangles - 40-50\%). Upper plots show the
$\phi$-averaged radii (from left to right: $R_{\rm out}$, $R_{\rm side}$,
$R_{\rm long}$, lower plots show the magnitude of the allowed oscillations
(from left to right $R^{2}_{\rm out,2}$, $R^{2}_{\rm side,2}$,
$R_{\rm out-side,2}$), all as functions of the pair transverse-momentum $k_{T}$.
\label{fig:alicevskt}}
\end{figure*}

In Fig.~\ref{fig:resultsrhictolhc} we show an example comparison of
the $k_{T}$ dependence of the radii at RHIC and LHC energies at one
of the centralities (in this case top 5\%). Two main features are
apparent: the overall increase of the system size as well as larger
gradient (steeper slope) of transverse radii considered as functions of
$k_T$. The latter is a consequence of the larger averaged transverse flow
developed at the LHC. This is illustrated in Fig.~\ref{fig:hs} -
compare the numbers indicating transverse velocity at freeze-out on
panels b) and c). The cause for the former was already discussed,
now we concentrate on one particular feature. While the overall size of
the system is indeed larger at the LHC, the radii do not seem to grow in
the same way. The $out$ radius grows significantly less that the other
two, which is best illustrated by the $R_{\rm out}/R_{\rm side}$
ratio, which decreases from $1.1$ at RHIC to $0.95$ at the LHC. It
comes from a qualitative change in the results of the hydrodynamic
calculation seen in Fig.~\ref{fig:hs}. At RHIC the hypersurface at 
freeze-out temperatures was ``outside-in'', or in other words particles
at the larger transverse distances froze out earlier. We refer the
reader to our previous work~\cite{Kisiel:2006is} where this effect was
studied in the simplified form by analyzing the modified Blast-wave
parametrization. The RHIC data were found to be consistent with the so-called
``negative a'' scenario, that is the ``outside-in'' freeze-out, which
is also consistent with our more detailed hydrodynamic calculation,
and other hydrodynamic calculations. On the
other hand the so-called ``positive a'', or ``inside-out'' scenario was
inconsistent with the data mainly due to too small $R_{\rm out}$. The
detailed hydrodynamics calculation for LHC energies shown in this work exhibit
a qualitative change in the freeze-out shape to the ``inside-out''
type. Therefore we do expect to see a smaller, or more precisely, less
increased outwards radii, and that is exactly what we see in
Fig.~\ref{fig:resultsrhictolhc}. The effect is large,
when compared with the expected experimental systematic uncertainties and
therefore can be easily tested. Its confirmation in the data would be
a very strong indication of the existence of the hydrodynamically
behaving medium in relativistic heavy-ion collisions, and would indicate that we
are significantly advanced in understanding the dynamics of this medium.

\subsection{LHC results as function of $k_T$}

\begin{figure*}[tb]
\begin{center}
\includegraphics[angle=0,width=0.9 \textwidth]{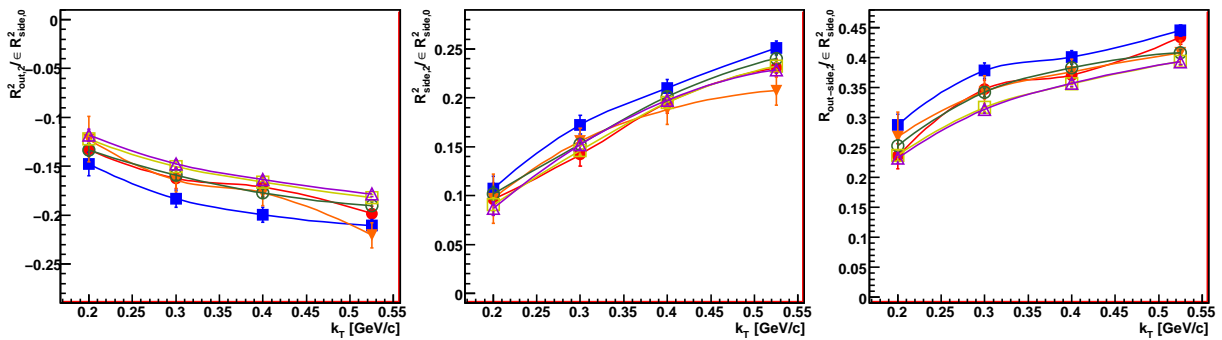}
\end{center}
\vspace{-6.5mm}
\caption{(Color online) Scaled azHBT oscillations for the LHC initial conditions for 6
centralities (filled circles - 0-5\%, filled squares -
5-10\%, down triangles 10-20\%, open circles -
20-30\%, open squares - 30-40\%, up triangles - 40-50\%). 
The symbols show the magnitude of the allowed oscillations 
(from left to right: $R^{2}_{\rm out,2}$, $R^{2}_{\rm side,2}$,
$R_{\rm out-side,2}$) divided by the initial eccentricity $\epsilon$ from
Table~\ref{tab:ab} and by $R_{\rm side}^2$.
\label{fig:oscvsepsilon}}
\end{figure*}

In Fig.~\ref{fig:alicevskt} we show the azimuthally sensitive HBT
results for the LHC energies as a function of the pair transverse momentum
$k_{T}$, including the low momentum bin of $0.05 - 0.15$~GeV. The
$out$ and $side$ radii fall with $k_{T}$ in a similar way,
linearly. The expected decrease of the size with growing centrality is
also apparent, while the $R_{out}/R_{side}$ ratio is close to a constant for all
centralities and values of $k_{T}$. The $long$ radius also falls
as expected. In general, the trends in the radii, plotted both versus
centrality and $k_{T}$, are smooth and well understood. It is interesting to
note the behavior of the oscillations. Their sign is the same as for LHC as for
RHIC. Although we do see a decrease in the magnitude of oscillations
with lowering $k_{T}$, especially in the $side$ direction, they do not
change sign. Also the magnitude of the oscillations grows consistently with
increasing centrality for all $k_{T}$ bins. 

The oscillations obtained from the analysis of the HBT correlation
functions correspond to the system asymmetry at the final stages of
the collision, when the freeze-out occurs. It is interesting to compare
them to the initial space asymmetry, as obtained from the Glauber
calculations and presented in Tab.~\ref{tab:ab}. This is shown in
Fig.~\ref{fig:oscvsepsilon}. One can see that after dividing the
observed asymmetry  by its initial value one obtains a curve which
appears to be universal, within the statistical uncertainties of this
study, for all considered centralities. 

\begin{figure*}[tb]
\begin{center}
\includegraphics[angle=0,width=0.99 \textwidth]{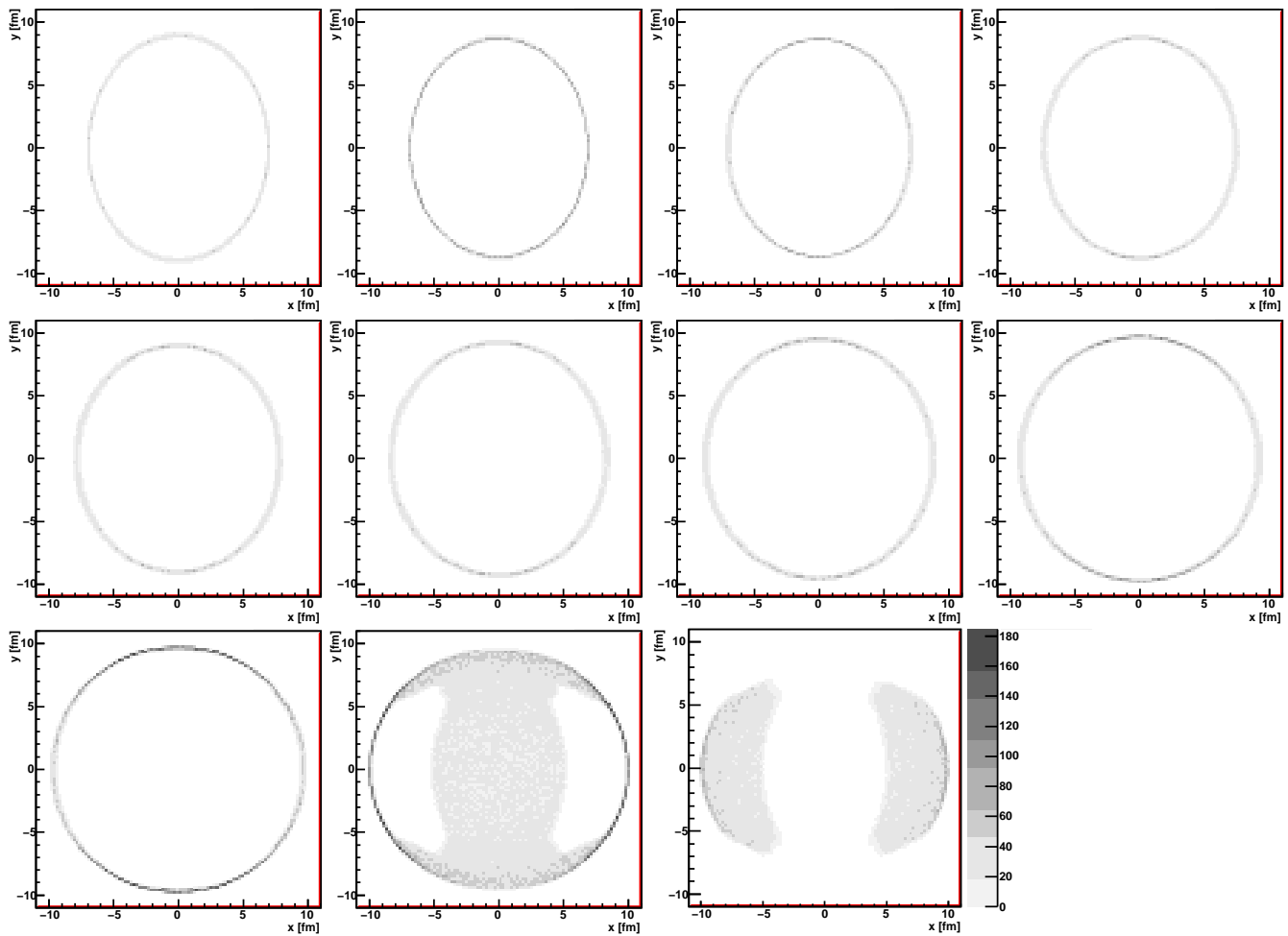}
\end{center}
\vspace{-6.5mm}
\caption{
Evolution of the shape of the system with time for the LHC energy and
centrality 20-30\%, as marked with the produced primordial pions. Each
panel shows the birth places of primordial charged pions emitted in a
time interval of duration 1~fm. The pions originate from the region
placed at the center of the collision ($ |z| <1 $ fm). The top-left
panel corresponds to the time interval 1.0-1.7 fm, top middle-left to
1.7-2.7 fm, and so on up to the time interval 10.7-11.7 fm shown in
the lower-right panel. The shades of gray indicate the relative number
of primordial charged pions emitted from the area element in the
transverse $(x,y)$ plane within the given time interval.  
\label{fig:lhcanatomy}}
\end{figure*}

\subsection{Emission history}

\begin{figure*}[tb]
\begin{center}
\includegraphics[angle=0,width=0.99 \textwidth]{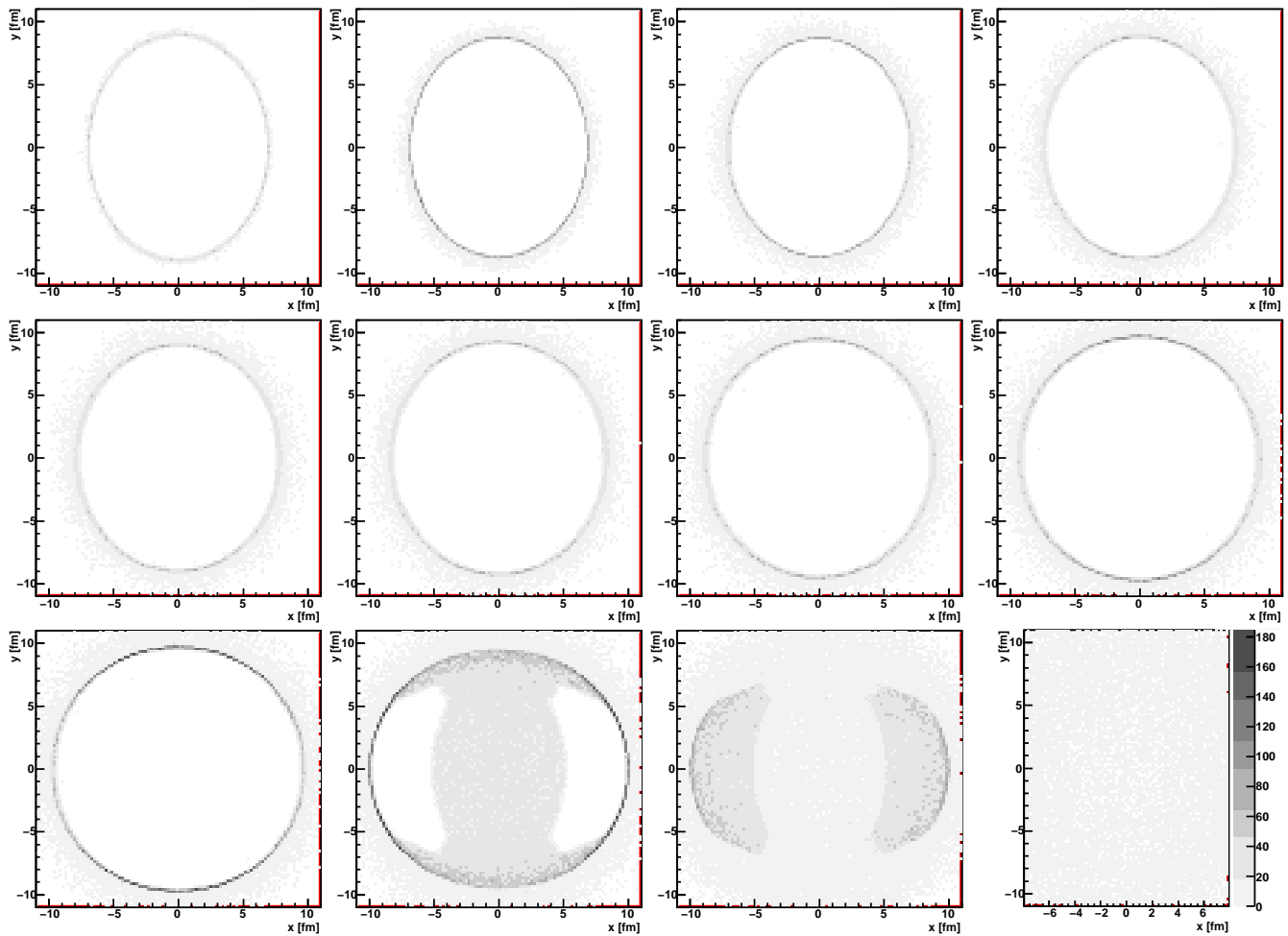}
\end{center}
\vspace{-6.5mm}
\caption{
Same as Fig.~\ref{fig:lhcanatomy} for all pions, primordial and from resonance decays. 
\label{fig:lhcanatomy_all}}
\end{figure*}

The above considerations constitute an overall picture that can be
deduced from the radii alone. However, since the particles are emitted
during the whole evolution time, the observed shape is essentially a
multiplicity-weighted average of the shape evolution of the system at
subsequent times. An example of such shape evolution with indicated
birth points of the primary pions is shown in
Fig.~\ref{fig:lhcanatomy} for the 20-30\% centrality for the PbPb
collisions at LHC.  The information discussed below is a more detailed
analysis of the behavior which may be inferred from  Fig.~\ref{fig:hs}.
The panels in Fig.~\ref{fig:lhcanatomy} show where
the charged primordial pions are emitted in a given time interval. The
scale (shown in the bottom-right panel) indicates how many pions are
emitted from a given area element in the transverse plane. The picture
shows the slice centered at midrapidity (the $z$ coordinate of the 
particle emission satisfies the condition $|z| < 1$ fm). At the very
beginning of the collision ($t<1.0$ fm) there are no
particles emitted. However, we observe particle emission at the very early
stage - already in the first time slice (1.0-1.7 fm). Obviously the
system has  only begun its evolution, hence the
shape is still very much out-of-plane extended, reflecting the initial
overlap geometry. We also see that up to the very last evolution stage
the emission is only from the surface of the system. As the evolution
progresses particles are emitted from more and more spherical shells,
up to the time of 9.7 fm. At this moment system reaches the
freeze-out temperature in its full volume and emission changes its
nature from the surface-like shell emission to the bulk emission from the whole
volume. It is especially interesting to look at panel 10 (9.7-10.7
fm). One can see that the volume emission starts in the center of
the system first, this is a very good illustration of the
``inside-out'' type of emission discussed earlier. What is also
interesting is that even though the overall shape of the system is
very much in-plane extended, the inner emitting part has an
out-of-plane shape. In the final stage of the evolution (panel 11:
10.7 - 11.7 fm) the volume emission continues and finally the outer
parts of the system emit particles. As can be seen from the picture
and the discussion, the emission patterns coupled to the shape
evolution of the system are quite complex. There is no way to tell,
a priori, what will be the overall system shape seen by femtoscopy. One
must perform a detailed simulation of all the evolution stages, which
must also include a realistic simulation of the number of particles
coming from each stage of the evolution. This is precisely what we
have done in this work.

In Fig.~\ref{fig:lhcanatomy_all} we show the analog of
Fig.~\ref{fig:lhcanatomy}, but with all pions, not only the primordial
ones. We note that the resonance decays ``wash out'' the production
regions, which is effectively increasing the size.
This is due to the transverse flow, which has a tendency to carry out
the resonance outward before if decays into pions. Qualitative
conclusions concerning the surface and volume emission are the same as
for Fig.~\ref{fig:lhcanatomy}. We also note that pion emission
continues well past the the time of 11.7 fm, as resonances with longer
lifetimes subsequently decay. The detailed discussion of the influence
of strongly decaying resonances on the HBT correlation functions and
extracted radii can be found in~\cite{Kisiel:2006is}.

\section{Conclusions}

The results presented in this paper show that it is possible to
achieve a uniform description of the RHIC soft-hadronic data including
azimuthally sensitive HBT radii. The initial space asymmetry of the
source, tuned earlier to reproduce the value of $v_2$, turns out to be
precisely such that the azimuthal dependence of $R^{2}_{\rm side,2}$ and
$R^{2}_{\rm out,2}$ is also very well described. This verifies good consistency
of our approach consisting of relativistic hydrodynamics and
statistical hadronization -- only the model prediction for the 
{\em out-side} radius is significantly different from the data, which requires 
explanation.  For the LHC energy we predict
the moderate increase of the HBT radii and the decrease of their
azimuthal  oscillations. Such effects can be naturally explained by
longer evolution times at LHC.  The space-time patterns of particle emission
were discussed in detail. They indicate  that the shape of the system
seen by the femtoscopic methods is an average of the complex and
varying in time shape of the emitting source. 

In summary, the results of our calculations for RHIC and LHC
conditions show several notable features. First of all the
calculations for RHIC show a remarkable agreement with the broad
spectrum of the soft physics data. We have concentrated on the
transverse dynamics of the source, and our model appears to properly
describe not only the momentum part of the observed phase space, but
also the space-time part. The unique feature of the model is the
proper description of the azimuthal asymmetry in the $side$ and $out$
directions, again both in the momentum and space-time, which has been
achieved before only in simplified and non-dynamical blast-wave
parametrizations~\cite{Retiere:2003kf} which neglect the important
contributions of the strongly decaying resonances, as well as the
surface  emission. The underlying hypothesis of our work is that the
system created in heavy-ion collisions at RHIC behaves as a single
piece of matter, and can be described by the hydrodynamic
equations. These equations include the state-of-the-art equation of
state which assumes that the matter above a certain critical
temperature is in a deconfined phase (Quark Gluon Plasma). In our
calculation the evolving system at RHIC spends a significant amount of
time in that phase, and its properties are essential in shaping up the
final observables. This work provides another crucial confirmation
that such a hypothesis is consistent with the experimental data. While
in itself it does not constitute a proof that the Quark Gluon Plasma
is indeed created at RHIC, we stress that any alternative explanation
must at least achieve a similar agreement with the experimental data
to be considered viable. 

The system created in heavy ion collisions at the LHC is predicted to
be in many ways similar to the one created at RHIC, at least in the
sense that it also is expected to spend a long time in the deconfined
phase. In fact, this time is predicted to be significantly larger than at RHIC, such that
its influence on the final observables may be more pronounced. Again, we
work under the assumption that hydrodynamics provides a good
description of the matter in such conditions, which allows us to provide
predictions for final state observables at these energies. We stress
that the underlying mechanisms of the model do not change at all
between the RHIC and the LHC energies, only a few of the external parameters, such as the
initial nucleon-nucleon cross-section or the initial temperature are
changed in a reasonable way. Nevertheless, we are able to identify
significant changes in the observables (with respect to RHIC) that can
be easily measures in the LHC experiments. In particular in this work,
which focuses on femtoscopy, we have identified two of them which are particularly 
sensitive: the
decrease of final observed anisotropy of the source (relative to $R_{\rm side}^2$), reflected in the
decrease of the oscillations in $R_{out}$ and $R_{side}$ radii, and the
change from the ``inside-out'' to the ``outside-in'' type of freeze-out,
reflected in the femtoscopic radii themselves and best illustrated as
the decrease of the $R_{out}/R_{side}$ ratio. These features are specific
enough such that they provide strict tests of the validity of the
hydrodynamic hypothesis. If observed, they will be a strong argument
that systems at RHIC and the LHC can indeed, at least in the soft
sector, be described by essentially the same physics principles. 

\bibliography{azHBT_wb7}

\end{document}